\documentclass[prl,twocolumn] {revtex4}
\usepackage{graphicx}
\usepackage{amsfonts}
\usepackage{amssymb}

\newcommand{\im}{\mathrm{Im}}
\newcommand{\re}{\mathrm{Re}\,}
\newcommand{\sgn}{\mathrm{sgn}}
\newcommand{\sqr}{{\scriptscriptstyle{\square}}}
\newcommand{\tgl}{{\scriptscriptstyle{\triangle}}}

\begin{document}
\author{Eytan Grosfeld and Ady Stern}
\affiliation{Department of Condensed Matter, Weizmann Institute of
Science, Rehovot 76100, Israel}
\title{Electronic transport in an array of quasi-particles in the $\nu=5/2$ non-abelian quantum Hall state}
\begin{abstract}
The Moore-Read Pfaffian $\nu=5/2$ quantum Hall state is a $p$-wave
super-conductor of composite fermions. Small deviations from
$\nu=5/2$ result in the formation of an array of vortices within
this super-conductor, each supporting a Majorana zero mode near
its core. Here we consider how tunneling between these cores is
reflected in the electronic response to an electric field of
non-zero wave vector $\bf q$ and frequency $\omega$. We find a
mechanism for dissipative transport at frequencies below the
$\nu=5/2$ gap, and calculate the ${\bf q},\omega$ dependence of
the dissipative conductivity. The contributions we find depend
exponentially on $|\nu-5/2|^{-1/2}$.
\end{abstract}
\maketitle

The $\nu=5/2$ fractional quantum Hall state is expected to be
characterized by quasi-particles obeying non-abelian statistics.
There are strong indications that this state is well described by
the Moore--Read Pfaffian wavefunction \cite{bib1}, which may be
formulated within composite-fermion theory (each electron is bound
to two flux quanta) as a $p$-wave superconductor of composite
fermions (CFs) at zero magnetic field. Excitations in this
superconductor are vortices carrying half a flux quantum and an
electric charge of $e/4$, and fermions created in twos by breaking
pairs with an appropriate energy gap \cite{bib2,bib221}. The
Bogoliubov-de-Gennes (BdG) equation describing the fermionic
excitations of a two-dimensional (2D) $p$-wave superconductor
admits zero-energy solutions in the presence of well separated
vortices, one solution near each vortex' core; These solutions are
Majorana fermions $\gamma$, satisfying $\gamma^\dag=\gamma$. As a
consequence, the ground state is degenerate; For $2N$ well
separated vortices, the ground state degeneracy is $2^N$. The
adiabatic interchange of two vortices induces a unitary
transformation within the subspace of degenerate ground states.
Two such transformations do not necessarily commute; Hence vortex
excitations obey non-Abelian statistics. A related spin model
showing similar non-abelian excitations was recently studied by
Kitaev \cite{bib20}.

Experimental support to the Moore-Read
theory is still needed. Relating the theory, and in particular
the non-abelian nature of the quasi-particles,  to measurable
observables, is a major theoretical challenge. Interference
experiments may be a venue towards that goal
\cite{bib21,bib22,bib23}.

In this work we pursue a different method to probe the ground
state degeneracy as well as some of the properties of the Majorana
excitations, by considering the response of a quantum Hall system
near filling factor $\nu=5/2$ to an
external electric field of wave-vector $\bf q$ and frequency $\omega$. In a fractional quantum Hall system at a
filling factor $\nu=5/2\pm\varepsilon$ ($\varepsilon\ll 1$), the
density deviation from $\nu=5/2$ is accommodated by means of
quasi-particles (vortices) whose density is $8\varepsilon n$,
where $n$ is the density of electrons. For a perfectly clean
system, these quasi-particles form a lattice, and when their
density is large enough, tunneling between their cores should be
taken into account. The degeneracy of the ground state is
partially removed by this tunneling, and a band is formed with a
width of the order of the tunneling strength. The tunneling also breaks the particle-hole symmetry of the localized $\gamma_i$'s.

We study the electronic transport through that band for square and
triangular lattices. We find that due to the existence of the
band, there is a dissipative part to the conductivity below the
$\nu=5/2$ energy gap, with a unique ${\bf q},\omega$ dependence.
This contribution to the conductivity, which does not involve a
motion of the vortices, depends exponentially on
$|\varepsilon|^{-1/2}$, due to its origin in tunneling. There is a
qualitative difference between the two lattice types. The square
lattice is described by an effective massless Dirac Hamiltonian,
while the triangular one shows a gap of a fraction of the band
width. We calculate the dissipative part of the conductivity of
the CFs using Kubo's formula \cite{bib10}, and then map it to the
electronic conductivity by a Chern-Simon transformation
\cite{bib3}
$(\sigma^e)^{-1}=(\sigma^{cf})^{-1}+\frac{2h}{e^2}\hat{\epsilon}$,
(with $\hat{\epsilon}$ being the anti-symmetric tensor).  For the
square lattice, we find that the longitudinal and transverse CF
conductivities are respectively
\begin{eqnarray}
    \label{eq:CF-conductivity1}
    \re(\sigma_{\sqr,\|}^{cf},\sigma_{\sqr,\perp}^{cf})=\frac{e^2}{\hbar}\frac{\vartheta^2(a
    q)^2}{16}\left(\frac{|\omega|}{\eta_\sqr^{1/2}},\frac{3\eta_\sqr^{1/2}}{|\omega|}\right)\theta\left(\eta_\sqr\right)
\end{eqnarray}
where $\eta_{\sqr}=\omega^2-v_0^2 q^2$. Here $a$ is the lattice
constant, $v_0$ is the velocity characterizing the Dirac spectrum,
and $\vartheta$, to be defined below, is related to the tunneling
strength. For the triangular lattice, we find that the real part of
the conductivity is
\begin{eqnarray}
    \label{eq:CF-conductivity2}
    \re \sigma^{cf}_{\tgl,\|}=\re \sigma^{cf}_{\tgl,\perp}=\frac{e^2}{\hbar}\frac{\vartheta^2 (3 a q)^2 \eta_\tgl\theta(\eta_\tgl)}{8\left(|\hbar\omega|/\sqrt{3}t\right)}
\end{eqnarray}
where $\eta_\tgl=\frac{|\hbar\omega|}{\sqrt{3}t}-2-\frac{a^2
q^2}{4}$. As we explain below, the electronic conductivities are
suppressed by a factor of $\omega^2$ relative to the CF
conductivities.

There are four steps in the calculation leading to these response
functions. First, we specify the Hamiltonian describing the array.
This Hamiltonian turns out to be closely related to the
Azbel-Hofstadter (A--H) Hamiltonian \cite{bib31,bib32}, describing
electrons on a tight binding lattice in a magnetic field. Second,
we calculate the spectrum of the Hamiltonian. Third, we find how
the system couples to gauge fields by expressing the density
and current operators in terms of the Majorana
operators $\gamma_i$ (with $i$ the vortex index); we also present
a physical picture of this coupling. Finally, we calculate
the response functions.

Based solely on the requirement of Hermiticity and on the relation
$\gamma_i=\gamma_i^\dag$, a lattice of well separated vortices is
generally described by a tight binding Hamiltonian
\begin{eqnarray}
    \label{eq:maj-hamiltonian}
    H=i t\sum_{ij}s_{ij}\gamma_{i} \gamma_{j}
\end{eqnarray}
where $\gamma_i$ are the Majorana operators satisfying
$\{\gamma_i,\gamma_j\}=\delta_{ij}$, and where $i,j$ are
nearest-neighbors lattice site indices. The tunneling strength $t$
is real and positive. The matrix $s_{ij}=\pm$ is anti-symmetric
and indicates the sign of the tunneling along the bond $(i,j)$.
While the freedom to redefine $\gamma_i\to -\gamma_i$ makes the
elements $s_{ij}$ gauge dependent, the product of $s_{ij}$ over
bonds creating a closed path is gauge independent. We now show
that this product is determined by a non-trivial phase a Majorana
fermion accumulates when encircling a plaquette, and give a simple
formula for the effective flux per plaquette. This formula fixes the matrix $s_{ij}$ up to a choice of gauge.

In the absence of
tunneling between vortex cores, the localized solution to a 2D
$p$-wave BdG equation near a vortex embedded in a lattice of
vortices is given by
\begin{eqnarray}
    \label{eq:maj-spinor}
    \chi_{i}(\mathbf{r})=\left(\begin{array}{c} e^{-i\pi/4+\frac{i}{2}\int_{\mathbf{P}_i}^\mathbf{r}\mathbf{\nabla}\Phi_i(\mathbf{l})\cdot d\mathbf{l}}g(\mathbf{r}-\mathbf{R}_i) \\ e^{i\pi/4-\frac{i}{2}\int_{\mathbf{P}_i}^\mathbf{r}\mathbf{\nabla}\Phi_i(\mathbf{l})\cdot d\mathbf{l}}g(\mathbf{r}-\mathbf{R}_i)\end{array}\right)
\end{eqnarray}
This is an approximate zero energy eigenstate of the first
quantized 2D $p$-wave Hamiltonian $H_{\mathrm{BdG}}$ (see
\cite{bib2},\cite{bib4}) of an order parameter
$\Delta_0(\mathbf{r})\exp{i \Omega(\mathbf{r};\{\mathbf{R}_i\})}$,
where $\mathbf{r}$ is the 2D-space coordinate and
$\{\mathbf{R}_i\}$ are the vortices' positions and the phase
$\Omega(\mathbf{r};\{\mathbf{R}_i\})$ has the property of
increasing by $2\pi$ around any closed path surrounding one vortex
(clockwise). The phase appearing in the solution
(\ref{eq:maj-spinor}) is given by
$\Phi_{i}(\mathbf{r};\{\mathbf{R}_i\})=\Omega(\mathbf{r};\{\mathbf{R}_i\})+\arg(\mathbf{r}-\mathbf{R}_i)$,
where the first term originates from the order parameter and the
second one originates from the $p_x+i p_y$ pairing, which induces
a relative particle-hole angular momentum. The point
$\mathbf{P}_i$ is arbitrarily chosen close to the vortex core. The
real wavefunction $g(r)$ is localized at the vortex core. The
tunneling matrix elements for nearest neighbors are purely
imaginary, and are given by $\pm i t$ where
$t=\left|\im\int_\mathbf{r}
\chi\left(\mathbf{r}-a\hat{x}\right)\left[H_{\mathrm{BdG}}(\mathbf{r})-H^{(0)}_{\mathrm{BdG}}(\mathbf{r})\right]\chi(\mathbf{r})\right|$
is the tunneling strength, and where $H^{(0)}_{\mathrm{BdG}}$ is
the Hamiltonian in the absence of tunneling, of which Eq.
(\ref{eq:maj-spinor}) is an exact zero energy eigenvector. For well separated vortices, $t$ decreases exponentially with $a\sim \varepsilon^{-1/2}$.

To determine the matrix elements $s_{ij}$, we consider a Majorana operator hopping between $n$ vortices
along a closed path that forms a polygon whose edges connect the
vortices. We show that there exists a non-trivial phase
related to this path, given by {\it half the sum of the interior
angles of the polygon}. The origin of this phase is in an
interplay between the phase of the order parameter and the
$p$-wave pairing. First we calculate the tunneling matrix elements
$<\chi_i|H_{\mathrm{BdG}}|\chi_j>=t_{ij} \exp i\psi_{ij}$, where
we use the tight-binding assumption to neglect the spatial
dependence of the phase and explicitly set
$\mathbf{r}=(\mathbf{R}_i+\mathbf{R}_j)/2\equiv \mathbf{C}_{ij}$.
For all bonds $t_{ij}=t$, while $\psi_{ij}$ is
given by
\begin{eqnarray}
    \nonumber \psi_{ij}=\frac{1}{2}\int_{\mathbf{P}_i}^{\mathbf{P}_j}\mathbf{\nabla} \Omega(\mathbf{l})\cdot d\mathbf{l}+\frac{1}{2}\int_{\mathbf{P}_i}^{\mathbf{C}_{ij}}\mathbf{\nabla}\arg(\mathbf{l}-\mathbf{R}_i)\cdot
    d\mathbf{l}\\
    -\frac{1}{2}\int_{\mathbf{P}_j}^{\mathbf{C}_{ij}}\mathbf{\nabla} \arg(\mathbf{l}-\mathbf{R}_j)\cdot d\mathbf{l}
\end{eqnarray}
The first term depends only on the order parameter; it measures
the change of the phase of the spinor due to vortices enclosed in
the path. The second and third terms are the contributions to the
phase due to the relative particle-hole angular momentum induced
by the $p_x+i p_y$ pairing; they measure changes in the direction
of the path. Considering $n$ tunneling events $t^n \exp
i\left[\psi_{i_1 i_2}+\psi_{i_2 i_3}+\dots+\psi_{i_n i_1}\right]$,
the total phase is given by
$\frac{1}{2}\oint\mathbf{\nabla}\Omega(\mathbf{l})\cdot
d\mathbf{l}+\frac{1}{2}\sum_{i=1}^n A_i$, where $A_i$ is the angle
subtended by the path with respect to the $i$-th vortex, positive
for anti-clockwise traversal. The first term gives a $\pi$-winding
for each vortex enclosed in the path. For each of these enclosed
vortices, $A_i$ is given by minus the exterior angle, which can be
written as $-(2\pi-I_i)$, where $I_i$ is the interior angle. For
all other vortices $A_i=I_i$. We therefore get $\frac{1}{2}\oint
\mathbf{\nabla}\Omega\cdot d\mathbf{l}+\frac{1}{2}\sum_{i=1}^n
A_i=\frac{1}{2}\sum_i I_i$, i.e. half the sum of interior angles
of the polygon. This result is independent of whether the core of a vortex on the path is inside or outside of the polygon: if
a path is deformed as to cross a core of a vortex, both the term
related to the order parameter and the relevant angle $A_i/2$
acquire an extra $\pi$, and these two contributions cancel each
other. For a general polygon of $n$ vortices we get a phase of
$\pi n/2-\pi$; Consequently, for a lattice whose plaquette is a
polygon of $n$ vortices we get $n/4-1/2$ flux quanta per
plaquette.

We note that for the A--H problem of tight-binding
electrons on the same lattice with the same flux per plaquette,
the Hamiltonian is
\begin{eqnarray}
    \label{eq:hof-hamiltonian}
    H^h=i t \sum_{ij}s_{ij}c_i^\dag c_j
\end{eqnarray}
The Hamiltonians (\ref{eq:maj-hamiltonian}) and
(\ref{eq:hof-hamiltonian}) share the same Harper's equation, their
spectra are identical, but they differ considerably in the way
they couple to gauge fields. Yet, there exist relations between
their response functions.

Our determination of the effective flux in
(\ref{eq:maj-hamiltonian}) singles out then a chain of A--H type
problems, one for each value of $n$, where the flux per plaquette
is determined by the geometry of the lattice. There is a
qualitative difference between the triangular lattice, with an odd
$n$, and the square lattice, with an even $n$; the former breaks
time reversal symmetry in the effective A--H problem while the
latter does not. The honeycomb lattice, for which $n=6$, was
considered in \cite{bib20}.

In the next step we calculate the spectrum and eigenvectors of the
Hamiltonian (\ref{eq:maj-hamiltonian}). After identifying the
flux per plaquette, we choose a gauge which complies with
it, commonly breaking translational symmetry. Translational
symmetry is restored by choosing a unit cell which contains an
integer multiple of the flux quantum. The sites of the unit cell
are numbered $z=1,\ldots,s$; In this way, we divide our lattice
into $s$ sublattices. We aim at finding an operator $\Gamma^\dag$
satisfying the equation $[H,\Gamma^\dag]=E\Gamma^\dag$. We expand
it in local site operators as $\Gamma^\dag=\sum_i \lambda_i
\gamma_i$ ending with the following equation
\begin{eqnarray}
    i t\sum_{j}s_{ij}\lambda_{j}=E \lambda_{i}
\end{eqnarray}
Using translational symmetry, the solution for each sublattice $z$
can be written as $\lambda_{i}=e^{i \mathbf{k}\cdot \mathbf{R}_i}
\lambda_{z(i)}(\mathbf{k})$, where $z(i)$ is the sublattice to
which the site $\mathbf{R}_i$ belongs. The equation for
$\mathbf{\lambda}_z$ is given by
\begin{eqnarray}
    \tilde{H}_{zz'}\lambda_{z'}\equiv i t\sum_{z'}\sum_{j\in z} s_{ij} e^{i \mathbf{k}\cdot (\mathbf{R}_j-\mathbf{R}_i)} \lambda_{z'}=E
    \lambda_z
\end{eqnarray}
where the site $i$ is an arbitrarily chosen lattice site that
belongs to the $z'$ sublattice. We denote the corresponding
eigenvectors of $\tilde{H}$ by
$\mathbf{\lambda}^{(\alpha)}(\mathbf{k})$. This results in the
following operators
\begin{eqnarray}
    \label{eq:operators}
    \Gamma^{(\alpha)\dag}_\mathbf{k}=\sum_i
    \lambda^{(\alpha)}_i(\mathbf{k})\gamma_i=\sum_{z=1}^{s}\lambda^{(\alpha)}_z(\mathbf{k})\sum_{i\in z}e^{i
    \mathbf{k}\cdot \mathbf{R}_i} \gamma_{i}
\end{eqnarray}
which obey the usual fermionic anti-commutation relations
$\{\Gamma^{(\alpha)\dag}_\mathbf{k},\Gamma^{(\beta)}_{\mathbf{k}'}\}=\delta_{\alpha\beta}\delta(\mathbf{k}-\mathbf{k}')$
and
$\{\Gamma^{(\alpha)\dag}_\mathbf{k},\Gamma^{(\beta)\dag}_{\mathbf{k}'}\}=0$
for positive energy modes. In terms of these operators the
Hamiltonian is diagonal and is given by
$H=\sum_{\mathbf{k}\alpha}E_{\mathbf{k}\alpha}\Gamma^{(\alpha)\dag}_{\mathbf{k}}
\Gamma^{(\alpha)}_{\mathbf{k}}$.

For the square lattice the A--H hamiltonian has half a quantum of
flux per plaquette. We choose a gauge for which $s_{ij}=+$ along
columns and has alternating signs between adjacent rows. Having
translational invariance in doubled lattice vectors, we may split
the lattice sites into four sublattices, numbered $z=1,\ldots,4$.
The Hamiltonian $H$ may be written in a $4\times 4$ matrix
notation as
\begin{eqnarray}
    \label{eq:lattice-hamiltonian}
    \tilde{H}_\sqr=2 t \sigma_x\otimes\tau_z \sin(a k_x)+2 t \sigma_x\otimes\tau_x \sin(a k_y)
\end{eqnarray}
In the limit $|k|a\rightarrow 0$ the Hamiltonian
(\ref{eq:lattice-hamiltonian}) has a doubly degenerate gapless
isotropic Dirac spectrum $\epsilon_{k\alpha}=\sgn(\alpha)v_0|k|$,
with $\alpha=+2,+1,-1,-2$ and the characteristic velocity
$v_0=2at$ \cite{bib9}. The eigenvectors of Eq.
(\ref{eq:lattice-hamiltonian}) are given by
\begin{eqnarray}
    \lambda^{(1)*}_{-\mathbf{k}}=\lambda^{(-2)}_{\mathbf{k}}=\lambda^{(2)}_{-\mathbf{k}}=\lambda^{(-1)*}_{\mathbf{k}}
    =\frac{\left(i e^{i\theta_k},e^{i\theta_k},-i,1\right)}{2}
\end{eqnarray}
where $e^{i\theta_k}=(k_x+i k_y)/|k|$.

For the triangular lattice the A--H Hamiltonian has a quarter of a
flux quantum per plaquette. The Hamiltonian in the sublattice
representation is given by
\begin{eqnarray}
    \label{eq:lattice-hamiltonian-triangular}
    \tilde{H}_\tgl=2 t \sum_{i=1}^{2}\xi_i\sin(\mathbf{a}_i\cdot
    \mathbf{k})+\xi_3\cos(\mathbf{a}_3\cdot \mathbf{k})
\end{eqnarray}
where $\xi_1=I\otimes\tau_x$, $\xi_2=\sigma_y\otimes\tau_y$,
$\xi_3=\sigma_y\otimes\tau_z$, and where $\mathbf{a}_1=(a
\hat{x}-\sqrt{3}a\hat{y})/2$, $\mathbf{a}_2=(a
\hat{x}+\sqrt{3}a\hat{y})/2$, $\mathbf{a}_3=a\hat{x}$ are
the three lattice directions. There is a doubly degenerate
spectrum indexed again by $\alpha$,  and the spectrum is \cite{bib5}
$\epsilon_{\mathbf{k}\alpha}=\sgn(\alpha)\epsilon_{\tgl,k}$, where
\begin{eqnarray}
    \epsilon_{\tgl,\mathbf{k}}=\sqrt{2}t\sqrt{3+\cos(2 a k_x)-2\cos(a
    k_x)\cos(\sqrt{3}a k_y)}
\end{eqnarray}
The spectrum is gapped, and there are two minima at
$\mathbf{k}_{0}=\left(\pm\pi/3a,0\right)$, around which the
spectrum is quadratic
$\epsilon_{\mathbf{k}_0+\mathbf{\kappa},\alpha}\simeq
\sgn(\alpha)\sqrt{3}t\left(1+\frac{1}{2}a^2 \kappa^2\right)$. The
eigenvectors of Eq. (\ref{eq:lattice-hamiltonian-triangular}) are
given by
\begin{eqnarray}
    && \lambda^{(1)}_{\mathbf{k}}=\lambda^{(-2)*}_{-\mathbf{k}}=\frac{1}{N_\mathbf{k}}\left(i B_{-\mathbf{k}},-i B_{-\mathbf{k}},1,1\right)\\
    &&
    \lambda^{(2)}_{\mathbf{k}}=\lambda^{(-1)*}_{-\mathbf{k}}=\frac{1}{N_\mathbf{k}}\left(-i
    B_\mathbf{k}, -i B_\mathbf{k},-1,1\right)
\end{eqnarray}
where $N_\mathbf{\mathbf{k}}$ is a normalization factor and
\begin{eqnarray}
B_\mathbf{k}=\frac{\epsilon_{\tgl}(k)/2t+\sin(\mathbf{a}_1\cdot
\mathbf{k})}{\cos(\mathbf{a}_3\cdot \mathbf{k})+i
\sin(\mathbf{a}_2\cdot \mathbf{k})}
\end{eqnarray}

The coupling of the Majorana states of the Hamiltonian
(\ref{eq:maj-hamiltonian}) to an electric field is very different
from that of the electrons in the A--H Hamiltonian
(\ref{eq:hof-hamiltonian}), due to the particle--hole symmetry of
the operators $\gamma_i$. While each Majorana state
(\ref{eq:maj-spinor}) is electrically neutral, when tunneling
between vortex cores is switched on, a non-zero density of charge
appears {\it between} the vortices. Projected to the subspace of
the Majorana states, the density operator may be written as
$\rho(\mathbf{r})=\sum_{ij}\rho_{ij}(\mathbf{r})$ where
\begin{eqnarray}
    \rho_{ij}(\mathbf{r})=i s_{ij}g(\mathbf{r}-\mathbf{R}_i)g(\mathbf{r}-\mathbf{R}_j)\gamma_i\gamma_j
\end{eqnarray}
The operator $i\gamma_i\gamma_j$ has two eigenvalues $\pm 1$ which
describe the sign of the charge mostly sitting at the center of
the bond; However, the operators $\rho_{ij}$ do not commute if
they share a common Majorana operator, and consequently one cannot
specify the charge at all bonds simultaneously. Furthermore, two
nearest-neighbour spinors $\chi_i$ and $\chi_j$ are exactly
orthogonal $<\chi_i|\chi_j>=0$ due to a $\pi$ phase difference
between the overlap of the particles and that of the holes;
however, they do support non-zero matrix elements of the charge
operator $<\chi_i|\sigma_z|\chi_j>=i s_{ij}\vartheta$, where
$\vartheta=\int_\mathbf{r} g(\mathbf{r}-a\hat{x})g(\mathbf{r})$.
Consequently, the excitation $\Gamma^{(\alpha)\dag}_k$ carries a
charge of $\vartheta \epsilon_{k\alpha}/t$.

Next we identify the current operator. At $q=0$, the current is
found using the identity
\begin{eqnarray}
    \label{eq:majorana-current}
    \mathbf{j}_{\mathbf {q}=0}=i[H,\mathbf{d}]=-i\vartheta
    t\sum_{ijl}s_{ij}s_{jl}\left(\frac{\mathbf{R}_l-\mathbf{R}_i}{2}\right)\gamma_i\gamma_l
\end{eqnarray}
where $\mathbf{d}=\int_\mathbf{r} \mathbf{r} \rho(\mathbf{r})$ is
the total dipole operator. The sum $\sum_j
s_{ij}s_{jl}\neq 0$ only for sites $i$ and $l$ separated by a doubled lattice vector. The current may be transformed to $k$-space
by inverting (\ref{eq:operators}) and substituting it into
(\ref{eq:majorana-current}). The $\mathbf{q}=0$ current is a
conserved quantity. To see that, we examine the commutator of
$\mathbf{j}_{\mathbf{q}=0}$ with the Hamiltonian
\begin{eqnarray}
    \nonumber [H,\mathbf{j}_{\mathbf{q}=0}]\propto \sum_{ijlm}s_{ij}s_{jl}s_{lm}\left(\frac{\mathbf{R}_i+\mathbf{R}_m}{2}-\frac{\mathbf{R}_j+\mathbf{R}_l}{2}\right)\gamma_i\gamma_m
\end{eqnarray}
which is described by paths composed of three consecutive bonds
connecting the vortices given by $i,j,l,m$. All paths give zero
contribution as they interfere destructively with paths formed by
starting from one of the ends and reversing the order of steps to
the other end.

At finite $q$ we find the current by calculating
$\rho(\mathbf{q})=\int_\mathbf{r} e^{i \mathbf{q}\cdot
\mathbf{r}}\rho(\mathbf{r})$ and using charge conservation. The
current operator is, in momentum space,
\begin{eqnarray}
    \label{eq:maj-current-k-space}
    \mathbf{j}(\mathbf{q})=\sum_{\mathbf{k}\alpha\beta} \tilde{e}_k \mathbf{v}_\mathbf{k} n^{\alpha\beta}_{\mathbf{k},\mathbf{q}}\Gamma^{(\alpha)\dag}_{\mathbf{k}+\mathbf{q}/2}\Gamma^{(\beta)}_{\mathbf{k}-\mathbf{q}/2}
\end{eqnarray}
where
$n^{\alpha\beta}_{\mathbf{k},\mathbf{q}}=\mathbf{\lambda}^{(\alpha)*}_{\mathbf{k}+\mathbf{q}/2}\cdot
    \mathbf{\lambda}^{(\beta)}_{\mathbf{k}-\mathbf{q}/2}$ are the density matrix elements of the associated A--H problem, and $\tilde{e}_k=e\vartheta
\epsilon_k/t$,
$\mathbf{v}_\mathbf{k}=\partial_\mathbf{k}\epsilon_k$ are the
charge and velocity of the quasi-particle respectively. For
comparison, the longitudinal component of the current in the
associated A--H Hamiltonian (\ref{eq:hof-hamiltonian}) is
\begin{eqnarray}
    \label{eq:hof-current-k-space}
    j^h_{\|}(\mathbf{q})=\sum_{\mathbf{k}\alpha\beta}J^h_{\alpha\beta}(\mathbf{k},\mathbf{q})c^{(\alpha)\dag}_{\mathbf{k}+\mathbf{q}/2}c^{(\beta)}_{\mathbf{k}-\mathbf{q}/2}
\end{eqnarray}
where for the relevant transitions
$J^h_{1,-1}=J^{h*}_{2,-2}=\frac{e v_0}{2}\left
(e^{i\theta_{{\mathbf{k}+\mathbf{q}/2}}}-e^{-i\theta_{{\mathbf{k}-\mathbf{q}/2}}}\right)$
for the square lattice and $J^h_{1,-2}=-J^{h}_{2,-1}=\frac{e a
t}{2\sqrt{2}}\left(3+3\sqrt{3}i\right)$ for the triangular lattice
near the bottom of the band. Over all, the matrix elements of the
current operator in the Majorana problem
(\ref{eq:maj-current-k-space}) are smaller by a factor of $qa$
relative to those of the A--H problem
(\ref{eq:hof-current-k-space}).

Having calculated the spectrum and identified the relevant
operators, the response functions of the array of Majorana states
is readily calculated employing the Kubo formula, with the results
given by Eqs. (\ref{eq:CF-conductivity1}),
(\ref{eq:CF-conductivity2}). This result affirms the existence of
dissipative conductivity, hence the flow of in-phase current, even
at frequencies below the $\nu=5/2$ energy gap.

Two steps need to be taken to transform the composite fermions conductivities (\ref{eq:CF-conductivity1}) and (\ref{eq:CF-conductivity2}) into the measurable electronic conductivity. First,
the imaginary part of the CF conductivity, $i\rho_s e^2/\omega$
(with $\rho_s$ being the superfluid density of the CFs),
originating from the superconductivity of the CF condensate,
should be added to the calculated real part. Second, the
Chern-Simon transformation should be used to transform the CF
conductivity into the electronic one. In the limit of small
$\omega$, these two steps result in
\begin{eqnarray}
\label{sigmasigma} \re\sigma^e_{\parallel}({\bf q},\omega)&=&\left
(\frac{\omega}{2h\rho_s
}\right )^2\re\sigma^{cf}_{\perp} \nonumber \\
\re\sigma^e_{\perp}({\bf q},\omega)&=&\left
(\frac{\omega}{2h\rho_s }\right )^2\re\sigma^{cf}_{\parallel}
\end{eqnarray}

At finite temperature, assuming $v_0 q\ll \omega$, the
conductivity satisfies
$\sigma[T]=\sigma[T=0]\sgn(\omega)\tanh\frac{\hbar\omega}{4 k_B
T}$.

Before closing, we note that the same methods may be used to find
the response functions of the associated A--H problems. For the
square lattice the conductivity is $\re
(\sigma^h_{\sqr,\|},\sigma^h_{\sqr,\bot})=
\frac{1}{8}\frac{e^2}{\hbar}\left(\frac{|\omega|}{\eta_\sqr^{1/2}},\frac{\eta_\sqr^{1/2}}{|\omega|}\right)\theta(\eta_\sqr)$.
The value of the conductivity at the $q\rightarrow 0$ limit is a
universal $e^2/8\hbar$ \cite{bib9}; the origin of the universality
lies in an exact cancellation of the dependence on $v_0$ due to
the linear density of states $\propto \omega/v_0^2$. The
dependence of the conductivity on temperature is $
    \sigma^h_\sqr=\frac{1}{8}\frac{e^2}{\hbar}\sgn\left(\omega\right) \tanh\frac{\hbar\omega}{4k_B
    T}$.
 For the triangular lattice
the
conductivity at the bottom of the band is again universal,
$\frac{3}{4}\frac{e^2}{\hbar}$.

In summary, we calculated the electronic response of an array of
immobile quasi-particles of the $\nu=5/2$ state to an electric
field of non-zero ${\bf q},\omega$, due to tunneling between the
Majorana modes at their cores. We found a contribution to the
dissipative conductivity, Eq. (\ref{sigmasigma}), that is of a
unique ${\bf q},\omega$ dependence, and a strong exponential
dependence on the deviation of $\nu$ from $5/2$. Our analysis
neglected disorder, which will be discussed elsewhere.

We acknowledge financial support from the US-Israel BSF (2002-238)
and the Minerva foundation. We thank A. Kitaev and Z. Tesanovic
for instructive discussions.

\end{document}